\documentclass[lettersize,journal]{IEEEtran}
\usepackage{amsmath,amsfonts}
\usepackage{algorithmic}
\usepackage{algorithm}
\usepackage{array}
\usepackage[caption=false,font=normalsize,labelfont=sf,textfont=sf]{subfig}
\usepackage{textcomp}
\usepackage{stfloats}
\usepackage{url}
\usepackage{verbatim}
\usepackage{graphicx}
\usepackage{cite}
\usepackage{upgreek}
\begin{document}

\title{Free-Running Waveguide-Integrated Single-Photon Avalanche Detectors for Visible Light}

\author{Aswin Alexander, Anirudh R. Ramaseshan, Soe M. Thar, Thomas Y. L. Ang, Jing Zhou, Alexander Ling, and Victor Leong

\thanks{Aswin Alexander, Anirudh R. Ramaseshan, Jing Zhou and Victor Leong are with the 
Institute of Materials Research and Engineering (IMRE), Agency for Science, Technology and Research (A*STAR),
2 Fusionopolis Way, Innovis \#08-03, Singapore 138634, Republic of Singapore. 
Anirudh R. Ramaseshan is also with the Quantum Innovation Centre (Q.InC), Agency for Science, Technology and Research (A*STAR), 2 Fusionopolis Way, Innovis \#08-03, Singapore 138634, Republic of Singapore. (email: victor\_leong@a-star.edu.sg)}
\thanks{Soe M. Thar and Alexander Ling are with the Centre for Quantum Technologies, National University of Singapore, 3 Science Drive 2, Singapore 117543, Republic of Singapore.}
\thanks{Thomas Y.L. Ang is with the Institute of High Performance Computing (IHPC), Agency for Science, Technology and Research (A*STAR), 1 Fusionopolis Way, \#16-16 Connexis North Tower, Singapore 138632, Republic of Singapore.}}



\maketitle

\begin{abstract}
Waveguide-integrated single-photon avalanche detectors (SPADs) are essential components of integrated photonics platforms for scalable extreme-low-light applications without the use of cryogenics.
Here, we demonstrate an integrated SPAD for visible light operating at room temperature in a free-running mode without gating.
The device is based on a doped silicon diode end-fire-coupled to a silicon nitride (SiN) photonic integrated circuit (PIC). 
We investigate a range of lateral and vertical doping profile designs,
and operate the devices with a simple current-mode passive quenching circuit. 
The optimal device is a laterally-doped \mbox{p-i-n$^+$} SPAD with 
a maximum photon detection efficiency (PDE) of \(1.95 \pm 0.32 \%\) for input light at 685\,nm wavelength,
when reverse-biased at an excess of 1.5\,V beyond the breakdown voltage of 15.0\,V. 
We identify promising avenues for improving device performance, 
which would enable such integrated SPADs to be an attractive choice for cutting-edge integrated photonics solutions in quantum technologies, low-light imaging, and high-speed communications at visible wavelengths.
\end{abstract}

\begin{IEEEkeywords}
Integrated photonics, avalanche photodetectors, single-photon detectors
\end{IEEEkeywords}

\section{Introduction}
Single-Photon Avalanche Photodetectors (SPADs) are silicon-based diode photodetectors 
capable of single-photon detection:
when reverse-biased beyond the breakdown voltage in the so-called Geiger mode,
the absorption of a single incident photon can trigger a macroscopic avalanche signal.
They have become indispensable for low-light applications, 
including quantum information processing~\cite{ndagano2020imaging,ceccarelli2021,keshavarzian20233,ma2016design}, 
biomedical imaging~\cite{ingargiola_48-spot_2017,bruschini2019}, 
optical communications~\cite{9714289,huang_2023}, 
and light detection and ranging (LiDAR)~\cite{li2022spad,villa_spads_2021}. 
With the rapid progress of integrated photonics platforms in recent years,
the successful integration of SPADs with photonic integrated circuits (PICs) would spur further breakthroughs in
miniaturizing complex optical systems and the implementation of cutting-edge quantum technologies. 

While the prospect of integrating superconducting photodetectors with PICs has gained prominence due to their very high 
photon detection efficiency (PDE) and low dark count rate (DCR)~\cite{wolff2020superconducting,chang2023nanowire,colangelo2024molybdenum}, 
the requirement for cryogenic cooling systems may hamper their deployment in the field.
In contrast, SPADs are capable of operating at or near room temperature, making them a more attractive option for applications that seek to avoid the use of bulky and cost-ineffective cryogenic systems~\cite{kuzmenko20203d,jiang2023long}. 
Moreover, SPAD fabrication is compatible with mature Complementary Metal-Oxide-Semiconductor (CMOS) foundry processes, providing a clear avenue to scalability and mass production~\cite{povzar2024ultra,gramuglia2021engineering,jiang2021time}.

Although several integrated SPADs in PICs have been reported for devices operating at infrared (IR) wavelengths~\cite{martinez2017single,wang2023high,acerbi2024monolithically}, 
this remains a nascent field.
For visible wavelengths, several reports have demonstrated  integrated avalanche photodetectors (APDs) operating in the linear regime above single-photon levels~\cite{yanikgonul2021,lin2022,9998065}.
An early work on integrated SPAD for visible wavelength was limited to gated-mode operation, achieving a PDE exceeding 6\%~\cite{govdeli2024room}. 

Although gated-mode operation can potentially enhance the PDE by mitigating the impact of high DCR or afterpulsing, 
and provide improved performance in low-light or noisy scenarios~\cite{lunghi2012advantages,Koziy_2021,yu2024recent},
it also has certain drawbacks. 
The duty cycles associated with the gating operation renders the SPAD inactive for long intervals, thus preventing continuous and uninterrupted photodetection. 
Moreover, the need for precise timing synchronization between the detector and a heralded or pulsed light source adds to the operational complexity~\cite{losev2021single,xu2023compact}.

In contrast to gating, free-running SPADs can operate continuously without requiring synchronization to an external gating signal. 
This allows for asynchronous photon detection, which is essential for applications involving some important quantum light sources,
such as spontaneous parametric down-conversion (SPDC) with a continuous-wave pump, which generates photons with random arrival times at the photodetectors~\cite{yan_ultra_2012,villar2020entanglement}. 
Additionally, free-running operation enables the measurement of high count rates as well as detection events occurring in quick succession, 
which is important for high-speed communications~\cite{wu_free-running_2023}, LiDAR~\cite{yu_fully_2017}, and time-resolved
spectroscopy~\cite{alayed_characterization_2018}.

Here, we report a demonstration of a free-running, waveguide-integrated Si-SPAD operating in the visible spectrum.
The devices are operated at room temperature with a simple passive quenching scheme, 
and are fabricated at a commercial photonics foundry on a conventional silicon nitride (SiN) platform,
pointing to their suitability for scalable fabrication and deployment.
We investigate a range of doping profiles of the silicon diode junctions to explore the optimal SPAD design,
and analyze our results to identify key improvements that can be made in subsequent development iterations.

\section{Methods}
\subsection{\label{sec:level2}SPAD Design}

Silicon (Si) SPADs are well-suited for applications at visible wavelengths
due to their highly efficient absorption up to 1100\,nm. 
We implement our SPADs on silicon slabs that are doped with p-n$^+$ or p-i-n$^+$ junctions, which are integrated with a SiN PIC. 
We note that while our previous simulation studies used a slightly different ridge waveguide geometry,
we adopt a Si slab geometry here to avoid potential issues arising from electric field concentrations at the edges of the rib waveguide~\cite{yanikgonul20182d,yanikgonul_simulation_2020}.
For optimal performance, the SPAD should maximize the probability that the charge carriers generated via the absorption of an incident photon can trigger a macroscopic avalanche, 
which then constitutes the single-photon detection signal.
As such, the incident optical mode and the doping profiles are important considerations of the integrated SPAD design.

\begin{figure}
    
    \includegraphics[width=\linewidth]{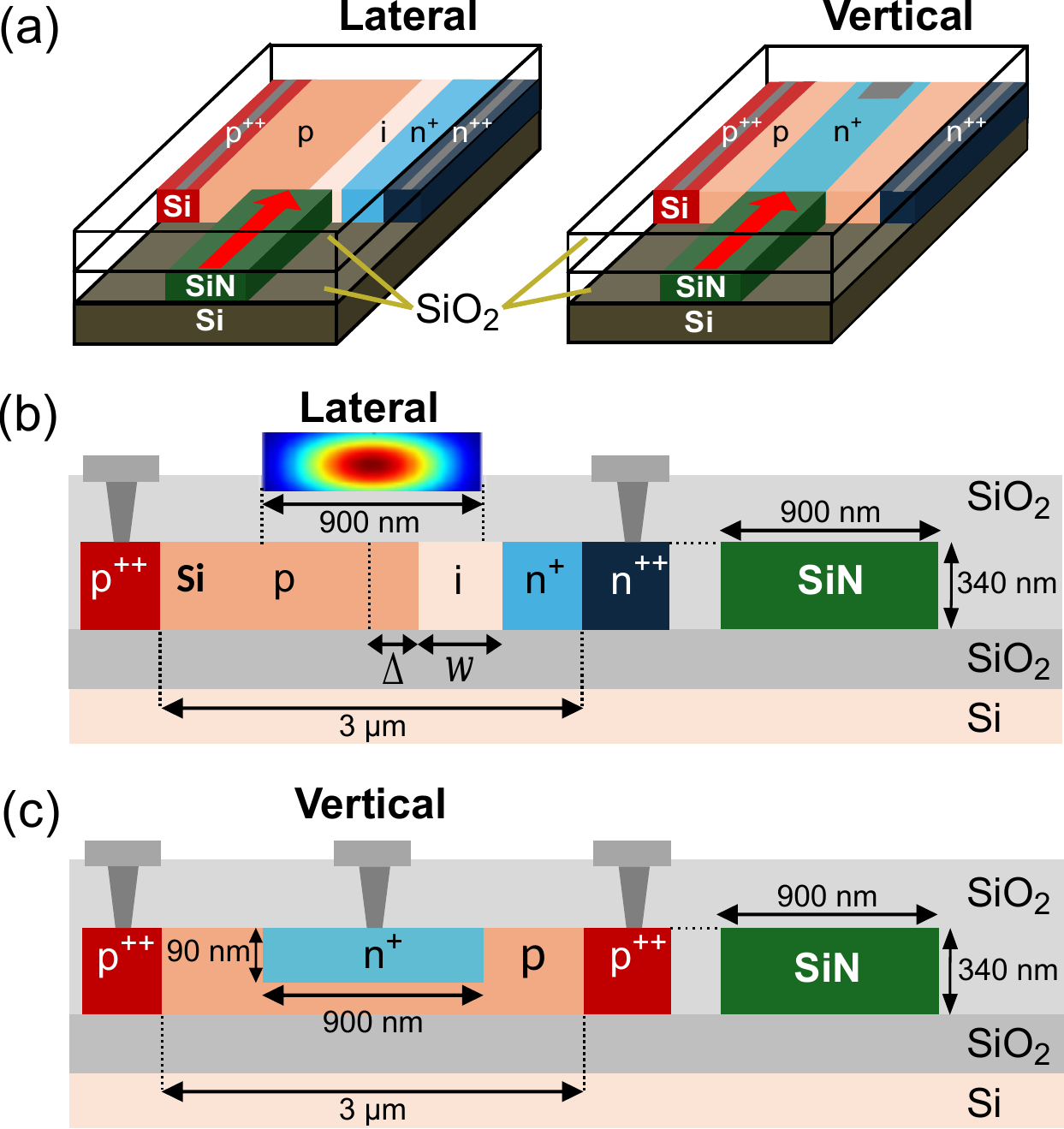}
    \caption{
    (a) Device schematics showing the geometry of the Si SPADs end-fire-coupled to the input SiN waveguide for both lateral (left) and vertical (right) doping profiles. 
    Gray regions denote the positions of the electrodes.
    The red arrow indicates the propagation direction of incident light. 
    (b),(c) Cross-section schematics of the lateral and vertical doping profiles, respectively. 
    The inset shows the optical mode profile at the SiN-Si interface.}
    \label{fig:spad_schematic}
\end{figure}

We adopt an end-fire coupling geometry where the Si SPAD and the SiN waveguide are on the same device layer (see Fig.~\ref{fig:spad_schematic}).
This allows for efficient coupling with a compact device footprint compared to a conventional interlayer coupling geometry, 
which requires long coupling lengths at visible wavelengths~\cite{lin2022}.
The Si slab and SiN waveguide have a thickness of 340\,nm,
and are cladded with 3\,$\upmu$m and 2\,$\upmu$m of SiO$_2$ above and below, respectively.
We investigate Si slab lengths of 9\,$\upmu$m and 16\,$\upmu$m, which would absorb $>$85\% of incident light at 685\,nm.

Based on the design optimization in our previous simulation studies~\cite{yanikgonul20182d,yanikgonul_simulation_2020}, 
the width of the SiN waveguide width at the Si-SiN interface is set at 900\,nm,
which is adiabatically widened from a single-mode SiN waveguide of width 400\,nm.
Thus, if the TE$_{00}$ mode of the single-mode waveguide is excited, 
the optical mode incident on the Si SPAD would also remain in the fundamental mode (see inset of Fig.~\ref{fig:spad_schematic}).
Inverse-taper edge couplers with a minimum width of 250 nm allow for efficient light coupling from lensed optical fibers to the SiN waveguides.

We investigate two different doping profiles types: lateral and vertical (see Fig.~\ref{fig:spad_schematic}(b), (c)),
where the descriptions refer to the predominant orientation of the electric field lines across the diode junction.
In this paper, we will refer to these devices as the lateral SPAD and vertical SPAD, respectively.

For the lateral SPAD, our simulations showed that the intrinsic layer width $w$ and displacement $\Delta$ from the center of the waveguide are important optimization parameters. 
Here, we investigate a total of 16 lateral p-i-n$^+$ junction variants across $\{w,\Delta\}=\{0,0.3,0.45,0.6\}$\,$\upmu$m. 
We note that an intrinsic width of $w$\,=\,0 is equivalent to a p-n$^+$ junction.

For the vertical SPAD, our study involves only one design, with the depth of the n\(^+\) region chosen to be 90 nm. 
Due to the charge diffusion effect and the fabrication resolution limits in defining the depth of the n$^+$ shallow-doped region,
it was not feasible to study multiple design variants for the vertical doping profile with our SPAD geometry.

The target doping concentration of the p and n$^+$ regions are \mbox{2$\times10^{17}$\,cm$^{-3}$} and \mbox{1$\times10^{19}$\,cm$^{-3}$}, respectively.
The lateral SPADs are electrically connected via metal pads and electrodes deposited on top 
of heavily doped p$^{++}$ and n$^{++}$ regions which are 3\,$\upmu$m apart,
and have a doping concentration of \mbox{1$\times10^{20}$\,cm$^{-3}$}. 
The vertical SPAD has an additional electrode at the end of the n$^+$ region.

\begin{figure*}[tb!]
    \centering
    \includegraphics[width=\linewidth]{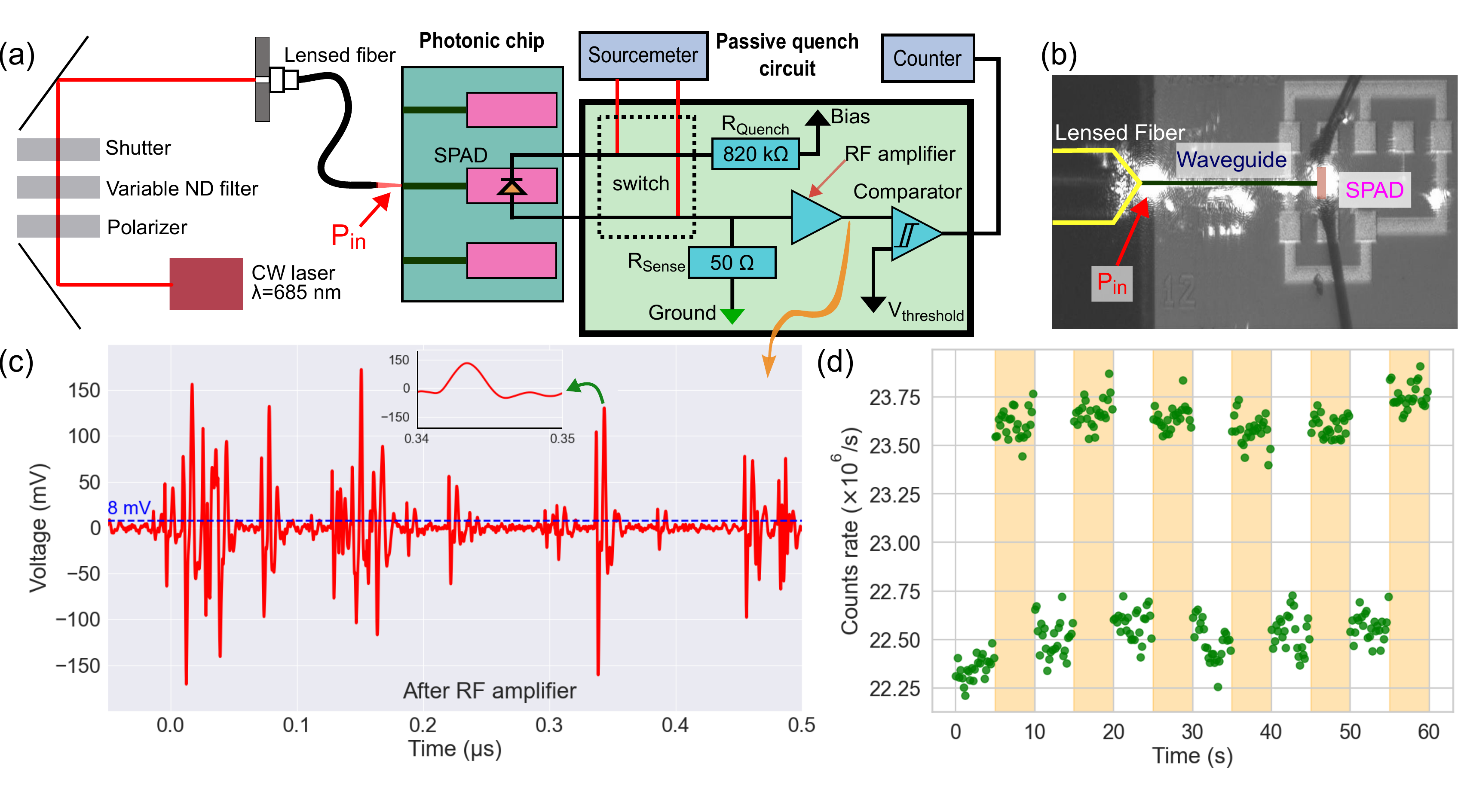}
    \caption{(a) Setup schematic. Input light is coupled via a polarization-maintaining lensed fiber to an on-chip photonic waveguide, and then to the waveguide-integrated SPAD. 
    The SPAD is wire-bonded to a passive quench circuit that generates output pulses based on the avalanche signals, which are then recorded by a counter device.  
    CW: continuous-wave, ND: neutral density.
    (b) Annotated optical micrograph of a wire-bonded SPAD device, with input light coupled from the lensed fiber to the on-chip waveguide. 
    (c) Avalanche signals from the SPAD after the RF amplifier. 
    The blue dotted line at 8 mV shows the comparator threshold used for identifying avalanche pulses. 
    The inset shows a magnified view of one of the pulses. 
    (d) Typical Geiger-mode signal of a SPAD with a lateral doping profile, subject to input light toggled by a shutter (yellow shaded region = on, white = off). 
    Input light power is $P_\mathrm{in}$\,=\,500\,pW. 
}
    \label{fig:setup}
\end{figure*}

\subsection{Passive Quenching Circuit}
When the SPAD operates in the Geiger mode and successfully detects an incident photon, 
the resulting avalanche current is self-sustaining,  
and must be quenched to reset the SPAD for subsequent photon detection. 
While active quenching circuits can enable better SPAD performance limits in terms of higher count rates and reduced afterpulsing~\cite{ceccarelli2021},
we adopt a passive current-mode quenching circuit\cite{cova_avalanche_1996} here for simplicity and ease of implementation (see Fig.~\ref{fig:setup}a).
The choice of the quenching circuit configuration is discussed in more detail in the Supplementary Information.

Briefly, a large ballast resistor $R_\mathrm{Quench}$ quenches the avalanche, 
while the output pulse across a sense resistor $R_\mathrm{Sense}$ serves as the SPAD signal.
We choose $R_\mathrm{Quench}$\,=\,820\,k$\Omega$: 
a smaller value results in a higher avalanche multiplication gain and hence larger output pulse amplitudes which are easier to detect,
but also yields a higher dark count rate as the noise (e.g. from thermal excitations) also experiences a larger amplification.
We choose $R_\mathrm{Sense}$\,=\,50\,$\Omega$ to avoid impedance mismatch with coaxial cables.

The SPAD signal (on the order of sub-millivolts) is amplified using a low-noise RF amplifier (Minicircuits ZKL-33ULN-S+)
and sent to a comparator (Texas Instruments TLV3502). 
The comparator threshold was set at 8\,mV to capture the majority of the SPAD pulses 
while rejecting electrical noise from the circuit (see Fig.~\ref{fig:setup}c.).
The 3\,V output pulses from the comparator are recorded by a counter (Keysight 53220a). 

In our implementation, the passive quench circuit board also contains a relay switch 
that allows the SPAD to be directly connected to an external instrument, such as a sourcemeter for current-voltage (I-V) measurements, 
instead of the quenching circuitry.

\subsection{Experimental Setup}

The SPADs were characterized at room temperature with the experimental setup shown in Fig.~\ref{fig:setup}a. 
The devices were wire-bonded directly to the quenching circuit to avoid a large stray capacitance associated with long electrical connections.
A 685\,nm continuous-wave laser (Thorlabs LP685-SF15) serves as the input light source.
A polarizer was used to optimize the coupling into a polarization-maintaining (PM) lensed fiber, 
which was oriented to couple to the fundamental transverse electric (TE) mode of the SiN waveguide.
A variable neutral density (ND) filter sets the desired input power $P_\mathrm{in}$,
and a shutter toggles the input light between ``on'' and ``off'' states.
The fiber-chip alignment was performed by motorized stages and optimized via an algorithm utilizing a feedback signal from the measured photocurrent from the SPAD device.
The entire setup was enclosed and verified to be light tight.

\begin{figure*}
    \centering
    \includegraphics[width=0.75\textwidth]{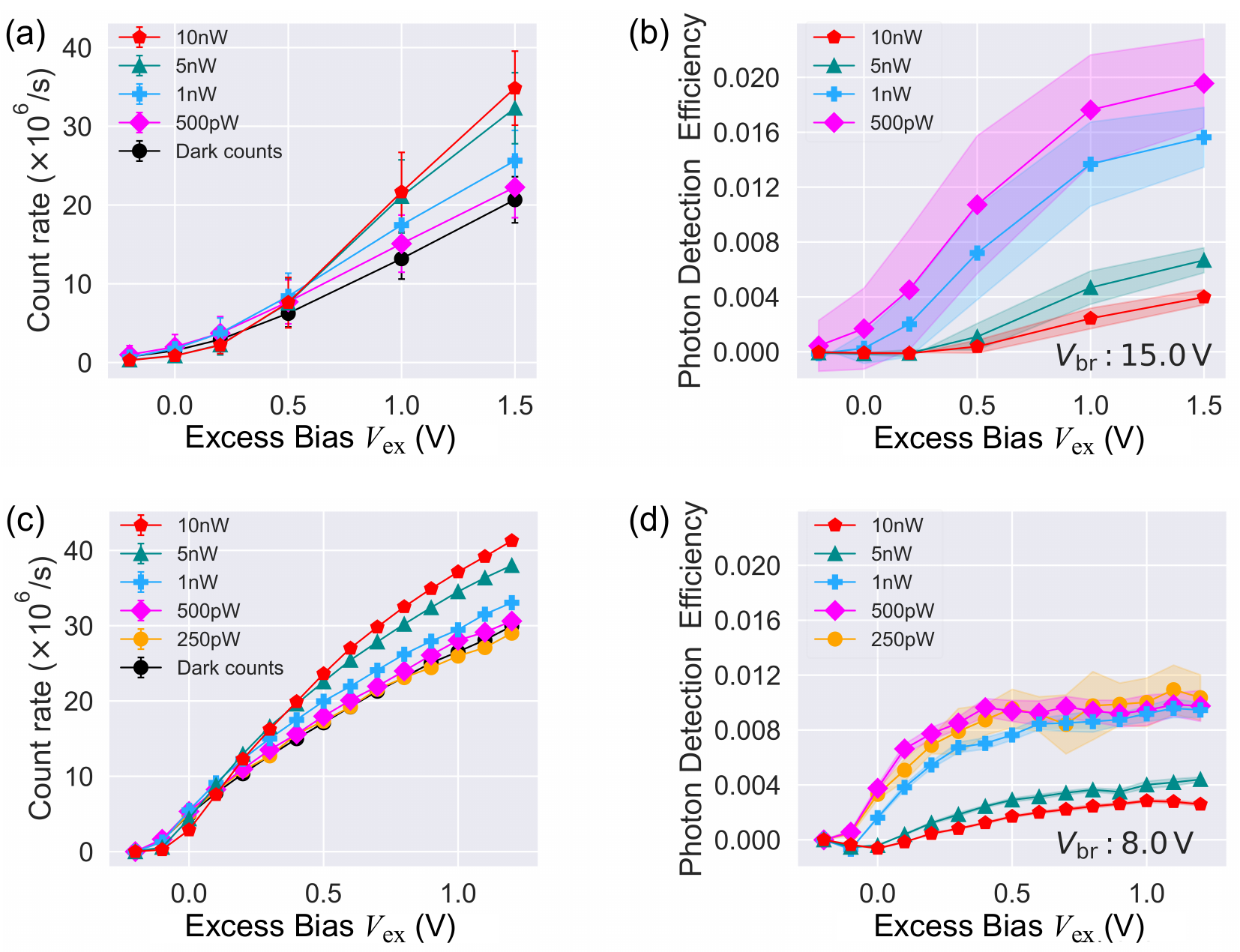}
    \caption{Measured SPAD performance at varying input optical powers $P_\mathrm{in}$ as a function of the excess bias $V_{\mathrm{ex}}$ above the breakdown voltage $V_\mathrm{br}$. 
    (a),(b) show the count rates and photon detections efficiencies (PDE) of a lateral SPAD with $w$\,=\,0 and $\Delta$\,=\,0.45$\upmu$m, respectively.
    (c),(d) show the count rates and PDE of a vertical SPAD, respectively.
    The error bars and shaded areas represent 1\,s.d. uncertainty.}
    \label{fig:pde}
\end{figure*}

\subsection{Measurement Procedure}
The SPAD was first characterized via current-voltage \mbox{(I-V)} measurements using a sourcemeter (Keithley 2636B) to obtain the breakdown voltage $V_\mathrm{br}$ of the device (more details are given in the Supplementary Information). 
The fiber-chip alignment was then performed by optimizing the photocurrent at zero applied bias with an input power of $P_\mathrm{in}$\,=\,1\,mW.
The optimized photocurrent $I_\mathrm{ref}$ is taken as a reference value.

Some of the devices were significantly affected by a drift in the breakdown voltage, a phenomenon also observed previously~\cite{yanikgonul2021,9998065}. 
To mitigate this, the DCR of each SPAD was monitored at a fixed bias voltage with the input light blocked with the shutter; 
devices with a DCR that varied by more than 50\% within one minute were discarded.
In addition, a forward bias of 1.5\,V was applied for 2\,s prior to each measurement to reset the SPAD to its original breakdown voltage.

For measurements with input light, the shutter was operated with a period of 10\,s and a duty cycle of 50\%. 
Each measurement was performed continuously over one minute with alternate bright and dark cycles, and repeated three times. 
Between each measurement, the fiber-chip alignment was verified by re-measuring $I_\mathrm{ref}$.
The alignment was re-optimized if $I_\mathrm{ref}$ has drifted by more than a few percent.

For the output pulse amplitude analysis, 
the output of the RF amplifier was connected directly to an oscilloscope. 
Each measurement duration was set at 2 minutes, and three separate readings were averaged to obtain the pulse amplitude distribution.

\section{Results and Discussion}

The devices were fabricated on a 8-inch wafers at a commercial photonics foundry (Advanced Micro Foundry) 
on a SiN-on-SOI platform.
The fabrication details closely follow our previous reports~\cite{yanikgonul2021,9998065}.
The photonic chips were subsequently wirebonded onto the passive quench circuit board.

A series of cutback measurements~\cite{yanikgonul2021} were carried out to determine the optical insertion loss of the devices, 
which consists of
the waveguide propagation loss ($0.270\pm0.005$\,dB/mm) over a length of 1.4\,mm, 
fiber-chip coupling loss ($5.30\pm0.18$\,dB per facet),
and the end-fire coupling loss at the Si-SiN interface ($3.95\pm0.07$\,dB per facet),
yielding a total insertion loss of $9.62\pm0.19$\,dB.
We note that these losses can be further mitigated with better fabrication control 
and optimized coupling structures for improved fiber-chip mode-matching, 
and our reported numbers do not represent the limits of the SiN platform.

From the DCR stability measurements, we decided to focus only on a subset of devices that exhibited more stable operation:
lateral and vertical doping profile devices of 16\,$\upmu$m and 9\,$\upmu$m length, respectively. 

\subsection{Photon Detection Efficiency (PDE)}

An important metric of a SPAD is the photon detection efficiency (PDE) $\eta$, 
which is the probability of a successful avalanche leading to the detection of an output pulse per input photon.
The PDE is given by:
\begin{equation}
\eta = \frac{n_{\text{light}} - n_{\text{dark}}}{n}
\label{eq:pde}
\end{equation}
where $n_{\text{light}}$ and $n_{\text{dark}}$ are the count rates of avalanche pulses detected when the shutter is opened and closed, respectively. 
Here, $n_{\text{dark}}$ is equivalent to the dark count rate (DCR).
The rate of input photons is:
\begin{equation}
n = \frac{\alpha \lambda P_{\text{in}}}{hc}
\label{eq:avg_photons}
\end{equation}
where $\alpha$ is the total insertion loss, $\lambda$ is the wavelength of the input light, $P_{\mathrm{in}}$ is the average optical power, $h$ is the Planck constant, and $c$ is the speed of light.

\begin{figure}
    \centering
    \includegraphics[width=\linewidth]{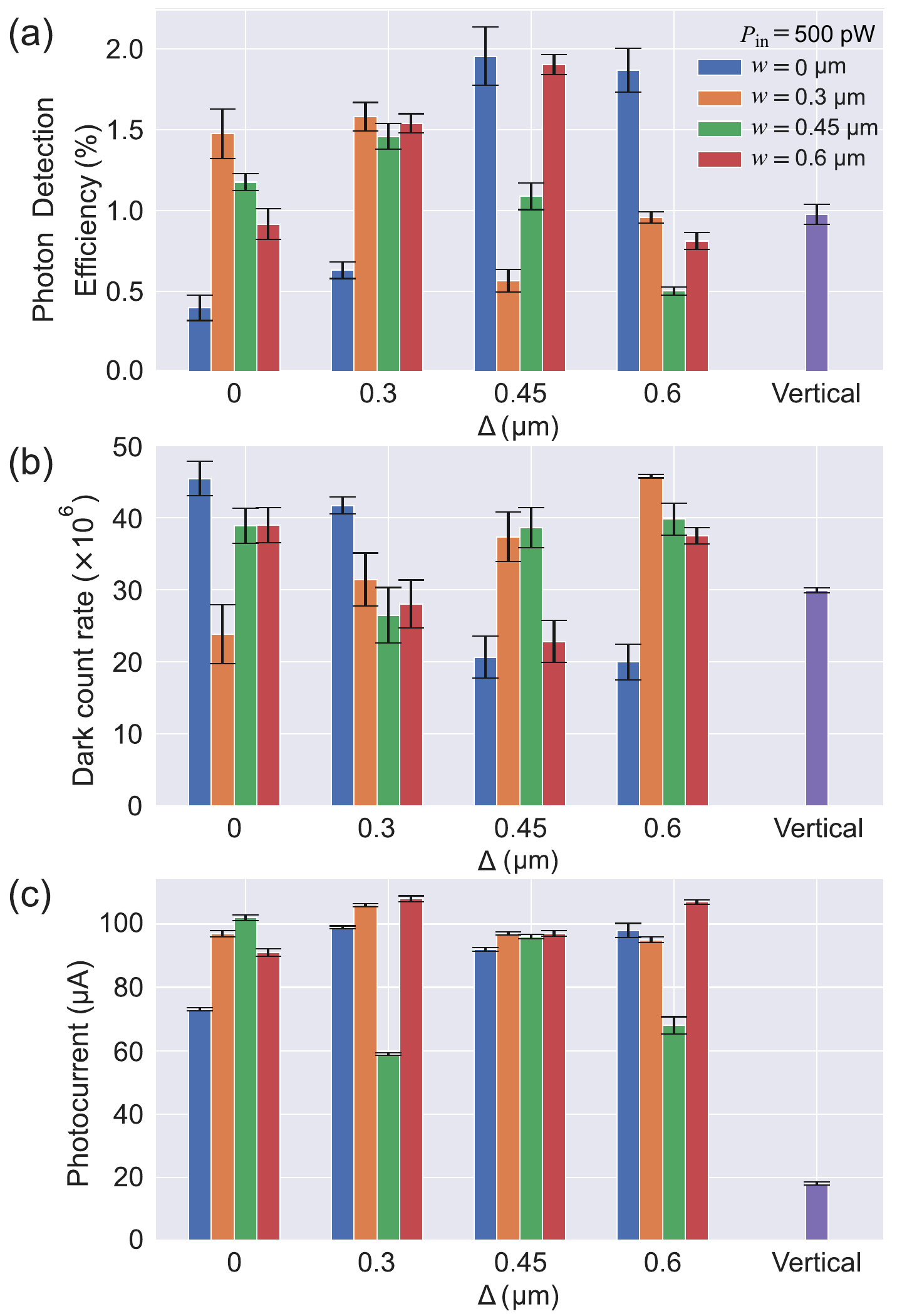}
    \caption{Comparison of SPAD performance across all doping profiles (16 lateral and 1 vertical).
    $\Delta$ and $w$ are the intrinsic region displacement and width, respectively, for the lateral doping profiles.
    (a)~Maximum PDE at input power $P_{\mathrm{in}}$\,=\,500\,pW,
    and excess bias $V_{\mathrm{ex}}$ of 1.5\,V and 1.2\,V for lateral and vertical doping profiles, respectively.  
    (b)~Dark count rate (DCR) for the same $V_{\mathrm{ex}}$ as in (a). 
    (c)~Linear-mode photocurrent measured at $P_{\mathrm{in}}$\,=\,1\,mW with zero applied bias.
    The maximum PDE has an inverse relationship with the DCR, but appears to be uncorrelated with the linear-mode photocurrent.
}
    \label{fig:histogram}
\end{figure}

The measured count rates and PDE of a representative lateral and vertical SPAD are shown in Fig.~\ref{fig:pde}
as a function of the excess bias $V_\mathrm{ex}$ beyond the breakdown voltage $V_\mathrm{br}$.
From these results, we can illustrate some general observations that were made across all device variants.
First, the dark count rate (DCR) and PDE are on the order of $10^7$\,cps (counts per second) and $\sim$1\%, respectively. 
The SPAD performance is significantly poorer than commercial non-PIC-integrated SPADs, 
in spite of how our integrated SPADs should arguably have a lower DCR than conventional counterparts 
due to its smaller active volume.
We hypothesize that the actual doping concentrations are higher than expected, 
resulting in Zener breakdown which contributes to the high DCR.
The high DCR would also have contributed to the low PDE due to a saturation effect, 
where the SPAD is unable to detect photons incident during the dead time of the avalanches triggered by dark noise.

Second, with increasing $V_\mathrm{ex}$, both $n_\mathrm{light}$ and $n_\mathrm{dark}$ increase 
as a higher electric field results in a stronger avalanche response.
However, the PDE shows a saturation behavior with higher $V_\mathrm{ex}$.
Coupled with an increasing DCR that would have limited the practical usefulness of the SPAD, 
we did not investigate beyond a $V_\mathrm{ex}$ of 1.5\,V for laterally doped and 1.2\,V for vertically doped devices.

Third, the measured PDE generally decreases as input optical power $P_\mathrm{in}$ increases, 
which also indicates a saturation behavior with respect to the input photon rate.
This can be expected as the count rate attributed to detected incident photons ($n_\mathrm{light}-n_\mathrm{dark}$) 
is high ($>$10$^6$\,cps for $V_\mathrm{ex}>$\,1\,V) at the measured $P_\mathrm{in}$ values;
at these count rates, conventional SPADs also typically require a nonlinear correction factor to compensate for the saturation effect at these count rates.

\subsection*{Benchmarking Across Devices}

Fig.~\ref{fig:histogram} shows the measured SPAD performance for all device variants at the highest excess voltage $V_\mathrm{ex}$ values of 1.5\,V and 1.2\,V for laterally and vertically doped devices, respectively. 
More comprehensive results measured at various $V_\mathrm{ex}$ are reported in the Supplementary Information.

The maximum PDE of $1.95\pm0.32\%$ is achieved with a laterally doped SPAD with $w=0$ and $\Delta=0.45$\,$\upmu$m, 
when biased at $V_\mathrm{ex}=1.5\,\mathrm{V}$ above its breakdown voltage of $V_\mathrm{br}=15.0$\,V. 
The vertically doped SPAD achieved a maximum PDE of $0.97\pm0.11\%$ when biased at $V_\mathrm{ex} = 1.2\,\mathrm{V}$ above its breakdown voltage of $V_\mathrm{br} = 8.0$\,V.

For the lateral p-n$^+$ junction devices (intrinsic width \mbox{$w=0$}), the PDE increases with the junction displacement~$\Delta$, with a slight drop for $\Delta=0.6$\,$\upmu$m.
This is consistent with our previous simulation results~\cite{yanikgonul_simulation_2020}
(though we note again the slight difference in waveguide geometries):
a larger~$\Delta$ increases the distance over which charge carriers can undergo avalanche multiplication, and thus the likelihood of a successful avalanche.
However, $\Delta$ being too large would also weaken the electric field in the SPAD, which then lowers avalanche efficiency. 
For the p-i-n$^+$ devices, no clear trend was identified in our results with respect to $w$ or $\Delta$, although previous simulations suggested PDE would increase for larger $w$, 
where a wider high-field region would boost avalanche multiplications. 

While we observe an inverse correlation between DCR and PDE (devices with higher PDE have lower DCR) 
due to saturation (see Fig.~\ref{fig:histogram}b),
we could not observe other trends that were predicted from simulations:
for instance, p-i-n$^+$ devices were expected to have have much lower DCR than p-n$^+$ devices~\cite{yanikgonul_simulation_2020}.
This is likely due an to unexpected mechanism (e.g. Zener breakdown) being the dominant source of dark counts in our devices.

We also analyzed the linear-mode photocurrent measured under zero applied bias 
for an input power of $P_{\mathrm{in}}$\,=\,1\,mW (see Fig.~\ref{fig:histogram}.c).
With a few exceptions, the measured photocurrents for the laterally doped SPADs were relatively uniform at $\sim$\,100\,$\upmu$A, 
corresponding to a responsivity of $\sim$\,0.5\,A/W.
Notably, the vertical SPAD had a much lower responsivity of $<0.1$\,A/W compared to the lateral devices, 
which cannot be fully accounted for by its shorter length (9\,$\upmu$m vs 16\,$\upmu$m).
The responsivity appears uncorrelated to the PDE or DCR, 
which suggests that the linear and Geiger mode performance of avalanche photodetectors are sensitive to different sets of device parameters.

\subsection{Pulse Amplitude Distribution}
\begin{figure*}
    \centering
    \includegraphics[width=0.9\linewidth]{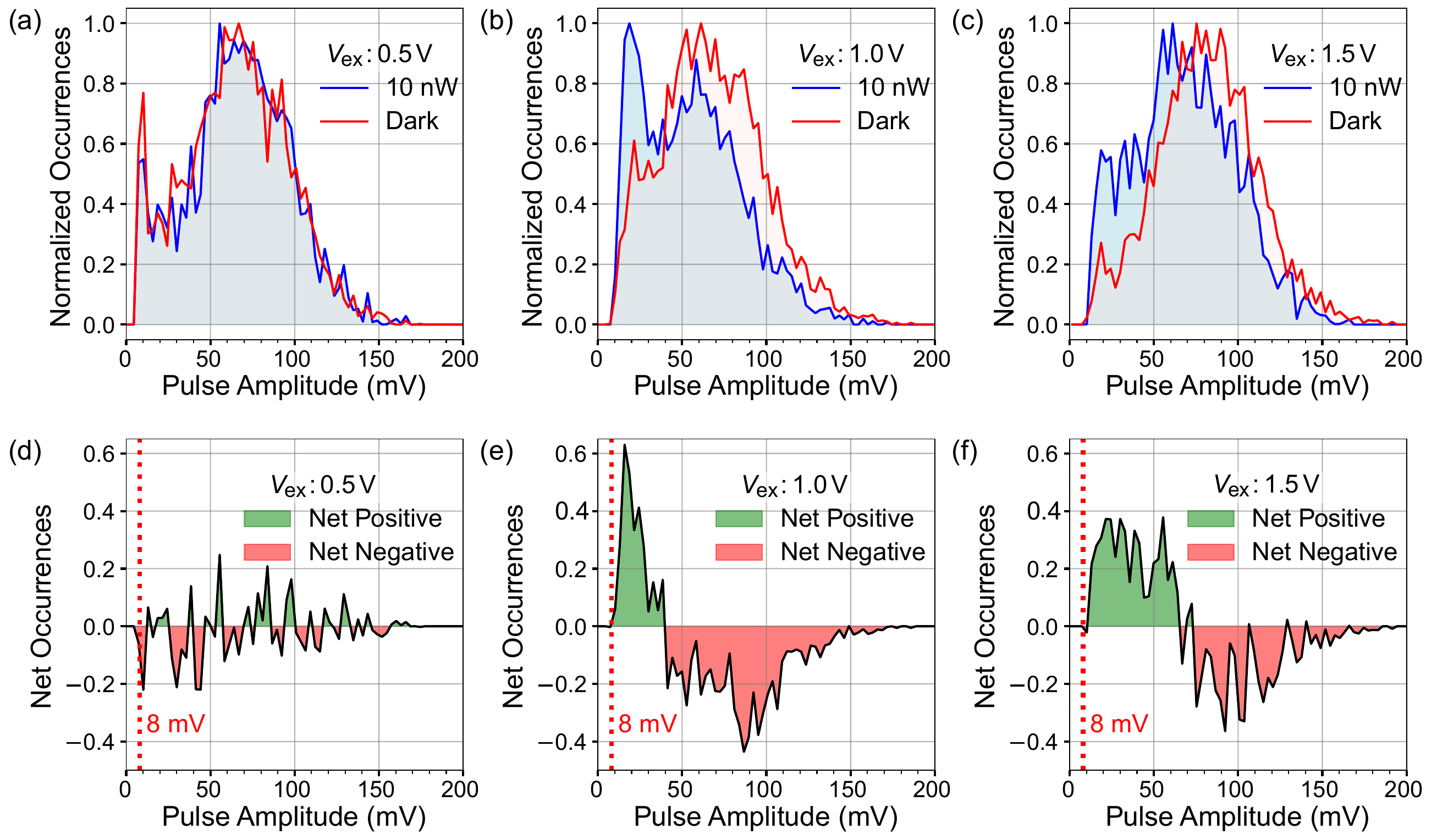}
    \caption{Distribution of the output pulse amplitudes after the RF amplifier for a lateral SPAD with $w$\,=\,0 and $\Delta$\,=\,0.  
    (a), (b), (c) show the normalized distributions at excess bias $V_\mathrm{ex}$ of 0.5 V, 1.0 V, 1.5 V, respectively,
    in dark conditions (red curve) and with input light at $P_{\mathrm{in}}$\,=\,10\,nW (blue curve).
    (d), (e), (f) show the corresponding net differences in the distributions, i.e. subtracting the red curve from the blue curve. 
    Here, a net positive occurrence at a particular pulse amplitude indicates a larger proportion of pulses of that amplitude occurring with input light compared to dark conditions.
    The 8\,mV comparator threshold is marked with red dotted lines.}
    \label{fig:pulse_dist}
\end{figure*}

To gain further insight to the device performance, we analyzed the SPAD pulses after the RF amplifier (example shown in Fig.~\ref{fig:setup}c).
The avalanche process in a SPAD is inherently stochastic 
due to the random nature of the impact ionization and avalanche multiplication processes; 
as well as the position of charge carrier generation due to photon absorption,
which is dependent on the optical mode within the SPAD structure~\cite{yanikgonul20182d}.
Moreover, in a high count rate regime, 
subsequent avalanches that are triggered before the SPAD is fully quenched and reset to the initial bias voltage
would yield smaller pulse amplitudes.
As such, a distribution of pulse amplitudes is expected.

Fig.~\ref{fig:pulse_dist} compares the typical pulse amplitude distribution for a SPAD at different excess bias voltages, in dark conditions and with input light at $P_{\mathrm{in}}$\,=\,10\,nW.
From the differences between the distributions with and without input light, 
we can identify the pulse amplitudes that are more strongly correlated with photogenerated avalanches (marked net positive in Fig.~\ref{fig:pulse_dist}) or with dark counts (net negative).
For higher excess bias $V_\mathrm{ex}$ of 1\,V and 1.5\,V, 
we can clearly identify that photogenerated avalanches account for a larger proportion of pulses with smaller amplitudes below 50\,mV and 70\,mV, respectively.
We deduce that the ability to simultaneously reject low-amplitude noise and large-amplitude pulses
could lead to the rejection of dark counts (hence reducing the DCR) without compromising the PDE.
However, our current quenching circuit only allows us to set a lower-bound threshold to reject low-amplitude noise,
and a more complex comparator scheme with this capability could be a meaningful direction for future work.

\section{Conclusion and Future Outlook}
In this work, we have shown a free-running waveguide-integrated SPADs for visible-light operation.
Devices with lateral and vertical doping profiles have obtained a maximum PDE of
$1.95\pm0.32$\% and $0.97\pm0.11$\%, respectively. 
The PDE is hampered by high dark count rates (DCRs) on the order of $10^7$\,cps, 
which we hypothesize to be the result of Zener breakdown;
future work will focus on fine-tuning the fabrication processes to mitigate or eliminate this effect.
Integrating these SPADs with active quenching circuits can further enhance their performance
by reducing the dead time and afterpulsing effects~\cite{cova1982active,lakeh2023integration}.
These integrated SPADs will broaden the applicability of integrated photonics platforms
in various fields such as photonic computing, quantum sensing, low-light imaging, and so on.

\section*{Acknowledgments}
We acknowledge the funding from the Agency for Science, Technology and Research, Singapore
(C222517002), and the National Research Foundation, Singapore, under the Quantum Engineering
 Programme (W21Qpd0304).
We also acknowledge the support from the National Quantum Federated Foundry (NQFF), Singapore
in the characterization of the photonic devices.


\bibliographystyle{IEEEtran}
\bibliography{refs}

@article{ndagano2020imaging,
  title={Imaging and certifying high-dimensional entanglement with a single-photon avalanche diode camera},
  author={Ndagano, Bienvenu and Defienne, Hugo and Lyons, Ashley and Starshynov, Ilya and Villa, Federica and Tisa, Simone and Faccio, Daniele},
  journal={npj Quantum Information},
  volume={6},
  number={1},
  pages={94},
  year={2020},
  publisher={Nature Publishing Group UK London}
}

@article{ceccarelli2021,
  title={Recent advances and future perspectives of single-photon avalanche diodes for quantum photonics applications},
  author={Ceccarelli, Francesco and Acconcia, Giulia and Gulinatti, Angelo and Ghioni, Massimo and Rech, Ivan and Osellame, Roberto},
  journal={Advanced Quantum Technologies},
  volume={4},
  number={2},
  pages={2000102},
  year={2021},
  publisher={Wiley Online Library}
}

@article{keshavarzian20233,
  title={A {3.3-Gb/s} {SPAD}-based quantum random number generator},
  author={Keshavarzian, Pouyan and Ramu, Karthick and Tang, Duy and Weill, Carlos and Gramuglia, Francesco and Tan, Shyue Seng and Tng, Michelle and Lim, Louis and Quek, Elgin and Mandich, Denis and others},
  journal={IEEE Journal of Solid-State Circuits},
  volume={58},
  number={9},
  pages={2632--2647},
  year={2023},
  publisher={IEEE}
}

@article{ma2016design,
  title={Design considerations of high-performance {InGaAs/InP} single-photon avalanche diodes for quantum key distribution},
  author={Ma, Jian and Bai, Bing and Wang, Liu-Jun and Tong, Cun-Zhu and Jin, Ge and Zhang, Jun and Pan, Jian-Wei},
  journal={Applied Optics},
  volume={55},
  number={27},
  pages={7497--7502},
  year={2016},
  publisher={Optica Publishing Group}
}

@article{ingargiola_48-spot_2017,
	title = {48-spot single-molecule {FRET} setup with periodic acceptor excitation},
	volume = {148},
	issn = {0021-9606},
	url = {https://doi.org/10.1063/1.5000742},
	number = {12},
	urldate = {2024-11-20},
	journal = {The Journal of Chemical Physics},
	author = {Ingargiola, Antonino and Segal, Maya and Gulinatti, Angelo and Rech, Ivan and Labanca, Ivan and Maccagnani, Piera and Ghioni, Massimo and Weiss, Shimon and Michalet, Xavier},
	month = nov,
	year = {2017},
}

@article{bruschini2019,
    author = {Bruschini, C. and Homulle, H. and Antolovic, I. M. and Burri, S. and Charbon, E.},
    title = {Single-photon avalanche diode imagers in biophotonics: review and outlook},
    journal = {Light: Sci. Appl.},
    volume = {8},
    pages = {87},
    year = {2019}
}

@ARTICLE{9714289,
  author={Huang, Shenjie and Safari, Majid},
  journal={IEEE Transactions on Communications}, 
  title={{SPAD}-Based Optical Wireless Communication With Signal Pre-Distortion and Noise Normalization}, 
  year={2022},
  volume={70},
  number={4},
  pages={2593-2605},
  doi={10.1109/TCOMM.2022.3151888}}

@article{huang_2023,
	title = {Performance analysis of {SPAD}-based optical wireless communication with {OFDM}},
	volume = {15},
	issn = {1943-0639},
	journal = {Journal of Optical Communications and Networking},
	author = {Huang, Shenjie and Li, Yichen and Chen, Cheng and Soltani, Mohammad Dehghani and Henderson, Robert and Safari, Majid and Haas, Harald},
	month = mar,
	year = {2023},
}

@article{li2022spad,
  title={{SPAD-based LiDAR} with real-time accuracy calibration and laser power regulation},
  author={Li, Dong and Hu, Jin and Ma, Rui and Wang, Xiayu and Liu, Yang and Zhu, Zhangming},
  journal={IEEE Transactions on Circuits and Systems II: Express Briefs},
  volume={70},
  number={2},
  pages={431--435},
  year={2022},
  publisher={IEEE}
}

@article{villa_spads_2021,
  title={SPADs and SiPMs arrays for long-range high-speed light detection and ranging (LiDAR)},
  author={Villa, Federica and Severini, Fabio and Madonini, Francesca and Zappa, Franco},
  journal={Sensors},
  volume={21},
  number={11},
  pages={3839},
  year={2021},
  publisher={MDPI}
}

@article{wolff2020superconducting,
  title={Superconducting nanowire single-photon detectors integrated with tantalum pentoxide waveguides},
  author={Wolff, Martin A and Vogel, Simon and Splitthoff, Lukas and Schuck, Carsten},
  journal={Scientific Reports},
  volume={10},
  number={1},
  pages={17170},
  year={2020},
  publisher={Nature Publishing Group UK London}
}

@article{kuzmenko20203d,
  title={{3D LIDAR} imaging using {Ge-on-Si} single-photon avalanche diode detectors},
  author={Kuzmenko, Kateryna and Vines, Peter and Halimi, Abderrahim and Collins, Robert J and Maccarone, Aurora and McCarthy, Aongus and Greener, Zo{\"e} M and Kirdoda, Jaros{\l}aw and Dumas, Derek CS and Llin, Lourdes Ferre and others},
  journal={Optics express},
  volume={28},
  number={2},
  pages={1330--1344},
  year={2020},
  publisher={Optica Publishing Group}
}

@article{povzar2024ultra,
  title={Ultra-Low Dark Count Rate {SPAD} Fully Integrated in a 180 nm High-Voltage {CMOS} Process},
  author={Po{\v{z}}ar, Borna and Berdalovi{\'c}, Ivan and Kne{\v{z}}evi{\'c}, Tihomir and Suligoj, Tomislav},
  journal={{IEEE} Photonics Technology Letters}, 
  year={2024},
  volume={36},
  number={20},
  pages={1241-1244}
}

@article{gramuglia2021engineering,
  title={Engineering breakdown probability profile for {PDP} and {DCR} optimization in a {SPAD} fabricated in a standard 55 nm {BCD} process},
  author={Gramuglia, Francesco and Keshavarzian, Pouyan and Kizilkan, Ekin and Bruschini, Claudio and Tan, Shyue Seng and Tng, Michelle and Quek, Elgin and Lee, Myung-Jae and Charbon, Edoardo},
  journal={IEEE Journal of Selected Topics in Quantum Electronics},
  volume={28},
  number={2: Optical Detectors},
  pages={1--10},
  year={2021},
  publisher={IEEE}
}

@article{jiang2021time,
  title={Time-gated and multi-junction {SPADs} in standard 65 nm {CMOS} technology},
  author={Jiang, Wei and Chalich, Yamn and Scott, Ryan and Deen, M Jamal},
  journal={IEEE Sensors Journal},
  volume={21},
  number={10},
  pages={12092--12103},
  year={2021},
  publisher={IEEE}
}

@article{yanikgonul2021,
    author = {Yanikgonul, Salih and Leong, Victor and Ong, Jun Rong and Hu, Ting and Png, Ching Eng and Krivitsky, Leonid},
    title = {Integrated avalanche photodetectors for visible light},
    journal = {Nature Communications},
    volume = {12},
    pages = {1834},
    year = {2021},
    doi = {10.1038/s41467-021-22046-x},
    url = {https://www.nature.com/articles/s41467-021-22046-x}
}

@article{lin2022,
    author = {Lin, Yiding and Yong, Zheng and Luo, Xianshu and Sharif Azadeh, Saeed and Mikkelsen, Jared and Sharma, Ankita and Chen, Hong and Mak, Jason C. C. and Lo, Patrick G. -Q. and Sacher, Wesley D. and Poon, Joyce K. S.},
    title = {Monolithically integrated, broadband, high-efficiency silicon nitride-on-silicon waveguide photodetectors in a visible-light integrated photonics platform},
    journal = {Nature Communications},
    volume = {13},
    pages = {6362},
    year = {2022},
    doi = {10.1038/s41467-022-34112-0},
    url = {https://www.nature.com/articles/s41467-022-34112-0}
}

@article{wang2023high,
  title={High-performance waveguide coupled Germanium-on-silicon single-photon avalanche diode with independently controllable absorption and multiplication},
  author={Wang, Heqing and Shi, Yang and Zuo, Yan and Yu, Yu and Lei, Lei and Zhang, Xinliang and Qian, Zhengfang},
  journal={Nanophotonics},
  volume={12},
  number={4},
  pages={705--714},
  year={2023},
  publisher={De Gruyter}
}

@article{govdeli2024room,
  title={Room-temperature waveguide-coupled silicon single-photon avalanche diodes},
  author={Govdeli, Alperen and Straguzzi, John N and Yong, Zheng and Lin, Yiding and Luo, Xianshu and Chua, Hongyao and Lo, Guo-Qiang and Sacher, Wesley D and Poon, Joyce KS},
  journal={npj Nanophotonics},
  volume={1},
  number={1},
  pages={2},
  year={2024},
  publisher={Nature Publishing Group UK London}
}

@article{Koziy_2021,
doi = {10.1070/QEL17566},
url = {https://dx.doi.org/10.1070/QEL17566},
year = {2021},
month = {aug},
publisher = {Kvantovaya Elektronika, Turpion Ltd and IOP Publishing},
volume = {51},
number = {8},
pages = {655},
author = {Koziy, A.A. and Losev, A.V. and Zavodilenko, V.V. and Kurochkin, Yu.V. and Gorbatsevich, A.A.},
title = {Modern methods of detecting single photons and their application in quantum communications},
journal = {Quantum Electronics}
}

@article{lunghi2012advantages,
  title={Advantages of gated silicon single-photon detectors},
  author={Lunghi, Tommaso and Pomarico, Enrico and Barreiro, Claudio and Stucki, Damien and Sanguinetti, Bruno and Zbinden, Hugo},
  journal={Applied Optics},
  volume={51},
  number={35},
  pages={8455--8459},
  year={2012},
  publisher={Optica Publishing Group}
}

@article{losev2021single,
  title={Single photon detectors based on SPADs: Circuit solutions and operating modes},
  author={Losev, AV and Zavodilenko, VV and Koziy, AA and Kurochkin, Yu V and Gorbatsevich, AA},
  journal={Russian Microelectronics},
  volume={50},
  pages={108--117},
  year={2021},
  publisher={Springer}
}

@article{xu2023compact,
  title={Compact free-running {InGaAs/InP} single-photon detector with 40\% detection efficiency and 2.3 kcps dark count rate},
  author={Xu, Qi and Yu, Chao and Chen, Wei and Zhao, Jianglin and Cui, Dajian and Zhang, Jun and Pan, Jian-Wei},
  journal={IEEE Journal of Selected Topics in Quantum Electronics},
  year={2023},
  volume={30},
  pages={1-7},
  publisher={IEEE}
}

@article{yan_ultra_2012,
	title = {An ultra low noise telecom wavelength free running single photon detector using negative feedback avalanche diode},
	volume = {83},
	issn = {0034-6748},
	url = {https://doi.org/10.1063/1.4732813},
	doi = {10.1063/1.4732813},
	urldate = {2024-12-03},
	journal = {Review of Scientific Instruments},
	author = {Yan, Zhizhong and Hamel, Deny R. and Heinrichs, Aimee K. and Jiang, Xudong and Itzler, Mark A. and Jennewein, Thomas},
	month = jul,
	year = {2012},
}

@article{wu_free-running_2023,
	title = {Free-Running Single-Photon Detection via {GHz} Gated {InGaAs}/{InP} {APD} for High Time Resolution and Count Rate up to 500 {Mcount}/s},
	volume = {14},
	issn = {2072-666X},
	urldate = {2024-12-03},
	journal = {Micromachines},
	author = {Wu, Wen and Shan, Xiao and Long, Yaoqiang and Ma, Jing and Huang, Kun and Yan, Ming and Liang, Yan and Zeng, Heping},
	month = feb,
	year = {2023},
        number={2},
        pages={437},
}

@article{yu_fully_2017,
	title = {Fully integrated free-running {InGaAs}/{InP} single-photon detector for accurate lidar applications},
	volume = {25},
	copyright = {\&\#169; 2017 Optical Society of America},
	issn = {1094-4087},
	url = {https://opg.optica.org/oe/abstract.cfm?uri=oe-25-13-14611},
	doi = {10.1364/OE.25.014611},
	journal = {Optics Express},
	author = {Yu, Chao and Shangguan, Mingjia and Xia, Haiyun and Zhang, Jun and Dou, Xiankang and Pan, Jian-Wei},
	month = jun,
	year = {2017},
}

@article{alayed_characterization_2018,
  title={Characterization of a time-resolved diffuse optical spectroscopy prototype using low-cost, compact single photon avalanche detectors for tissue optics applications},
  author={Alayed, Mrwan and Palubiak, Darek P and Deen, M Jamal},
  journal={Sensors},
  volume={18},
  number={11},
  pages={3680},
  year={2018},
  publisher={MDPI}
}

@article{acerbi2024monolithically,
  title={Monolithically integrated SiON photonic circuit and silicon single-photon detectors for NIR-range operation},
  author={Acerbi, Fabio and Bernard, Martino and Goll, Bernhard and Gola, Alberto and Zimmermann, Horst and Pucker, Georg and Ghulinyan, Mher},
  journal={Journal of Lightwave Technology},
  volume={42},
  number={8},
  pages={2831--2841},
  year={2024}
}

@ARTICLE{9998065,
  title={Visible-light integrated {PIN} avalanche photodetectors with high responsivity and bandwidth},
  author={Gundlapalli, Prithvi and Leong, Victor and Ong, Jun Rong and Ang, Thomas YL and Yanikgonul, Salih and Siew, Shawn Yohanes and Png, Ching Eng and Krivitsky, Leonid},
  journal={Journal of Lightwave Technology},
  volume={41},
  number={8},
  pages={2443--2450},
  year={2022},
  publisher={IEEE}
}

@article{yanikgonul_simulation_2020,
  title={Simulation of silicon waveguide single-photon avalanche detectors for integrated quantum photonics},
  author={Yanikgonul, Salih and Leong, Victor and Ong, Jun Rong and Png, Ching Eng and Krivitsky, Leonid},
  journal={IEEE Journal of Selected Topics in Quantum Electronics},
  volume={26},
  number={2},
  pages={1--8},
  year={2019},
  publisher={IEEE}
}

@article{cova_avalanche_1996,
  title={Avalanche photodiodes and quenching circuits for single-photon detection},
  author={Cova, Sergio and Ghioni, Massimo and Lacaita, Andrea and Samori, Carlo and Zappa, Franco},
  journal={Applied Optics},
  volume={35},
  number={12},
  pages={1956--1976},
  year={1996},
  publisher={Optica Publishing Group}
}

@article{chang2023nanowire,
  title={Nanowire-based integrated photonics for quantum information and quantum sensing},
  author={Chang, Jin and Gao, Jun and Esmaeil Zadeh, Iman and Elshaari, Ali W and Zwiller, Val},
  journal={Nanophotonics},
  volume={12},
  number={3},
  pages={339--358},
  year={2023},
  publisher={De Gruyter}
}

@article{colangelo2024molybdenum,
  title={Molybdenum Silicide Superconducting Nanowire Single-Photon Detectors on Lithium Niobate Waveguides},
  author={Colangelo, Marco and Zhu, Di and Shao, Linbo and Holzgrafe, Jeffrey and Batson, Emma K and Desiatov, Boris and Medeiros, Owen and Yeung, Matthew and Loncar, Marko and Berggren, Karl K},
  journal={ACS Photonics},
  volume={11},
  number={2},
  pages={356--361},
  year={2024},
  publisher={ACS Publications}
}

@article{jiang2023long,
  title={Long range {3D} imaging through atmospheric obscurants using array-based single-photon {LiDAR}},
  author={Jiang, Peng-Yu and Li, Zheng-Ping and Ye, Wen-Long and Hong, Yu and Dai, Chen and Huang, Xin and Xi, Shui-Qing and Lu, Jie and Cui, Da-Jian and Cao, Yuan and others},
  journal={Optics Express},
  volume={31},
  number={10},
  pages={16054--16066},
  year={2023},
  publisher={Optica Publishing Group}
}

@article{martinez2017single,
  title={Single photon detection in a waveguide-coupled {Ge-on-Si} lateral avalanche photodiode},
  author={Martinez, Nicholas JD and Gehl, Michael and Derose, Christopher T and Starbuck, Andrew L and Pomerene, Andrew T and Lentine, Anthony L and Trotter, Douglas C and Davids, Paul S},
  journal={Optics express},
  volume={25},
  number={14},
  pages={16130--16139},
  year={2017},
  publisher={Optica Publishing Group}
}

@article{villar2020entanglement,
  title={Entanglement demonstration on board a nano-satellite},
  author={Villar, Aitor and Lohrmann, Alexander and Bai, Xueliang and Vergoossen, Tom and Bedington, Robert and Perumangatt, Chithrabhanu and Lim, Huai Ying and Islam, Tanvirul and Reezwana, Ayesha and Tang, Zhongkan and others},
  journal={Optica},
  volume={7},
  number={7},
  pages={734--737},
  year={2020},
  publisher={Optica Publishing Group}
}

@article{yu2024recent,
  title={Recent advances in {InGaAs/InP} single-photon detectors},
  author={Yu, Chao and Xu, Qi and Zhang, Jun},
  journal={Measurement Science and Technology},
  volume={35},
  number={12},
  pages={122003},
  year={2024},
  publisher={IOP Publishing}
}

@article{yanikgonul20182d,
  title={{2D} {M}onte {C}arlo simulation of a silicon waveguide-based single-photon avalanche diode for visible wavelengths},
  author={Yanikgonul, Salih and Leong, Victor and Ong, Jun Rong and Png, Ching Eng and Krivitsky, Leonid},
  journal={Optics Express},
  volume={26},
  number={12},
  pages={15232--15246},
  year={2018},
  publisher={Optica Publishing Group}
}

@article{cova1982active,
  title={Active-quenching and gating circuits for single-photon avalanche diodes {(SPADs)}},
  author={Cova, Sergio and Longoni, A and Ripamonti, Giancarlo},
  journal={IEEE Transactions on nuclear science},
  volume={29},
  number={1},
  pages={599--601},
  year={1982},
  publisher={IEEE}
}

@article{lakeh2023integration,
title={Integration of an ultra-fast active quenching circuit with a monolithic {3D} {SPAD} pixel in a 28 nm {FD-SOI} CMOS technology},
  author={Lakeh, Mohammadreza Dolatpoor and Kammerer, Jean-Baptiste and Schell, Jean-Baptiste and Issartel, Dylan and Gao, Shaochen and Cathelin, Andreia and Dartigues, Alexandre and Calmon, Francis and Uhring, Wilfried},
  journal={Sensors and Actuators A: Physical},
  volume={363},
  pages={114744},
  year={2023},
  publisher={Elsevier}
}


 




\vfill

\end{document}


\title{Supplementary Information: Free-Running Waveguide-Integrated Single-Photon Avalanche Detectors for Visible Light}

\author{Aswin Alexander}
\affiliation{Institute of Materials Research and Engineering (IMRE), Agency for Science, Technology and Research (A$^\ast$STAR), 2 Fusionopolis Way, Innovis \#08-03, Singapore 138634, Republic of Singapore}
\author{Anirudh. R. Ramaseshan}
\affiliation{Institute of Materials Research and Engineering (IMRE), Agency for Science, Technology and Research (A$^\ast$STAR), 2 Fusionopolis Way, Innovis \#08-03, Singapore 138634, Republic of Singapore}
\affiliation{Quantum Innovation Centre (Q.InC), Agency for Science, Technology and Research (A$^\ast$STAR), 2 Fusionopolis Way, Innovis \#08-03, Singapore 138634, Republic of Singapore}
\author{Soe M. Thar}
\affiliation{Centre for Quantum Technologies, 3 Science Drive 2, National University of Singapore, Singapore 117543, Republic of Singapore}
\author{Thomas Y. L. Ang}
\affiliation{Institute of High Performance Computing (IHPC), Agency for Science, Technology and Research (A$^\ast$STAR), 1 Fusionopolis Way, \#16-16 Connexis North Tower, Singapore 138632, Republic of Singapore}
\author{Jing Zhou}
\affiliation{Institute of Materials Research and Engineering (IMRE), Agency for Science, Technology and Research (A$^\ast$STAR), 2 Fusionopolis Way, Innovis \#08-03, Singapore 138634, Republic of Singapore}
\author{Alexander Ling}
\affiliation{Centre for Quantum Technologies, 3 Science Drive 2, National University of Singapore, Singapore 117543, Republic of Singapore}
\author{Victor Leong}
\email[]{victor_leong@a-star.edu.sg}
\affiliation{Institute of Materials Research and Engineering (IMRE), Agency for Science, Technology and Research (A$^\ast$STAR), 2 Fusionopolis Way, Innovis \#08-03, Singapore 138634, Republic of Singapore}


\maketitle

\section{Passive Quenching Circuit Configuration}

Passive quenching circuits can be configured to work in a voltage mode or current mode~\cite{cova_avalanche_1996}.  
In the voltage mode, the sense resistor is connected in series to the quench resistor on the same side of the diode, resulting in an output that is an attenuated replica of the voltage across the SPAD. 
This configuration also acts as an intrinsic low-pass filter, which is detrimental for fast pulses where the higher frequency components are important. 
We have observed that the voltage-mode configuration failed to produce a meaningful output
due to the high dark count rates resulting in temporally overlapped avalanche pulses.

Instead, we adopt a passive quench circuit with a current mode configuration, 
where the sense resistor is connected to the ground lead of the SPAD, while the quench resistor is on the opposite side (see Fig.~2, main text). 
This allows the current pulse generated by the SPAD to be directly measured across the sense resistor, offering the best performance for high count rates. 
The trade-off is that the output pulse amplitudes are significantly lower, 
making their detection more challenging.

\section{Current-Voltage Measurements}

\begin{figure}[H] 
    \centering
    \includegraphics[width=0.6\linewidth]{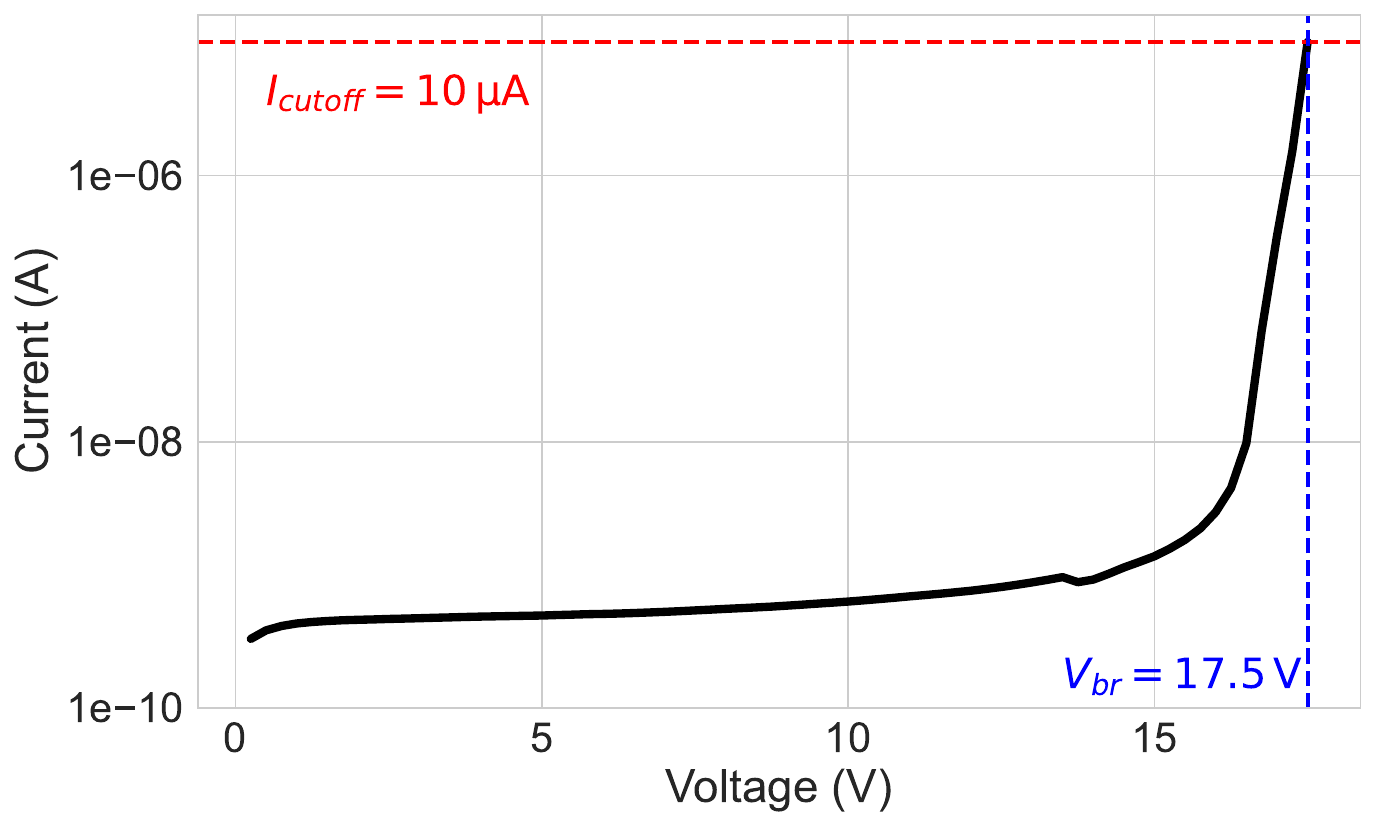}
    \caption{\label{fig:IV} Current-voltage (I-V) measurement of a p-i-n lateral junction SPAD 
    with an intrinsic region width $w$\,=\,0.6\,$\upmu$m and a junction displacement of $\Delta$\,=\,0.45\,$\upmu$m.
    The voltage sweep is performed in steps of 0.25\,V.
    The measurement is stopped once the breakdown voltage $V_\mathrm{br}$ is reached,
    which we define as the bias at which the current reaches a threshold of 10\,$\upmu$A.
    For this device, $V_\mathrm{br}$\,=\,17.5\,V.}
    \label{fig:i-v}
\end{figure}

\section{Afterpulsing}

\begin{figure}[H] 
    \includegraphics[width=0.85\linewidth]{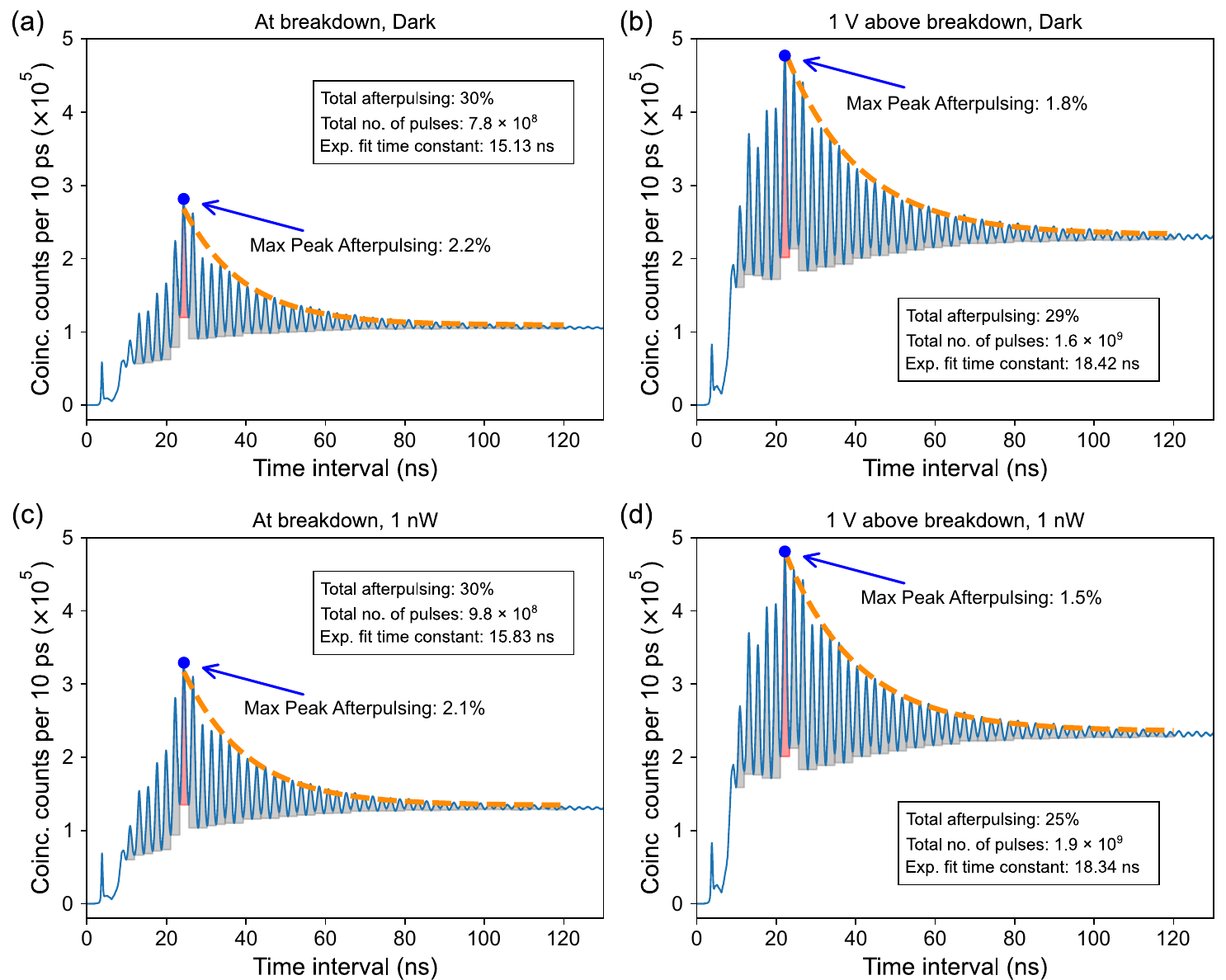}
    \caption{\label{fig:afterpulsing} 
    Auto-correlation histograms of the output pulses of a typical SPAD,
    showing the afterpulsing behaviour for different operating bias and input optical power.
    Each histogram is an average of several measurements of 60\,s each.
    The afterpulsing probability is calculated by summing the number of coincidence events under the peaks after subtracting the background 
    (represented by the shaded areas), and divided by the total number of SPAD pulses.
    We identify the max peak as the one with the highest number of counts (shaded in red), and the afterpulsing probability associated with that peak alone is labeled in the plots.
    The exponential fit is done based on the highest point of each peak, starting from the max peak.
    }
\end{figure}

To measure the afterpulsing effect, we analyzed the auto-correlation of the SPAD output pulses with a time tagger device (Swabian Instruments, Time Tagger Ultra).
The results for a typical SPAD are shown in Fig.~\ref{fig:afterpulsing} for different bias and input power conditions.
We expect that without afterpulsing, the auto-correlation histogram will show no coincidences at small time intervals within the dead time of the SPAD, which then rises to a flat background given by random dark noise;
afterpulsing would then show up as correlations above this background.

In each case, we observe the afterpulsing occuring as multiple peaks with a regular interval of $\sim$\,2.2\,ns. 
After the first few pulses, the peak amplitudes decay with an exponential time constant of 
$\sim$\,15\,ns and $\sim$\,18\,ns when biased at breakdown and 1\,V beyond breakdown, respectively.
To quantify the afterpulsing, we sum the number of events under the peaks after subtracting the background
(see shaded areas in Fig.~\ref{fig:afterpulsing}), 
and normalize it to the total number of SPAD pulses over the measurement window of 60\,s
to obtain an overall afterpulsing probability of 25--30\%.
As a reference, we also calculate the afterpulsing probability associated to only the largest peak, which gives a value of 1.5--2.2\%.

For higher bias, the background is higher (corresponding to a higher dark count rate), but does not seem to significantly affect the afterpulsing features or magnitude.
There are also no significant differences observed in the afterpulsing between the SPAD operated in dark conditions or with input light.
The periodic nature of the peaks suggest that it is closely linked to the underlying mechanism: 
if each afterpulsing event is caused by the trapping and subsequent release of charge carriers which then trigger the next avalanche,
then the peak intervals could reflect a characteristic time scale of the trap-release cycle.
Further investigation in future work could reveal more insights into the afterpulsing behaviour observed here.

\section{SPAD Performance}
\begin{figure}[H] 
    \includegraphics[width=\linewidth]{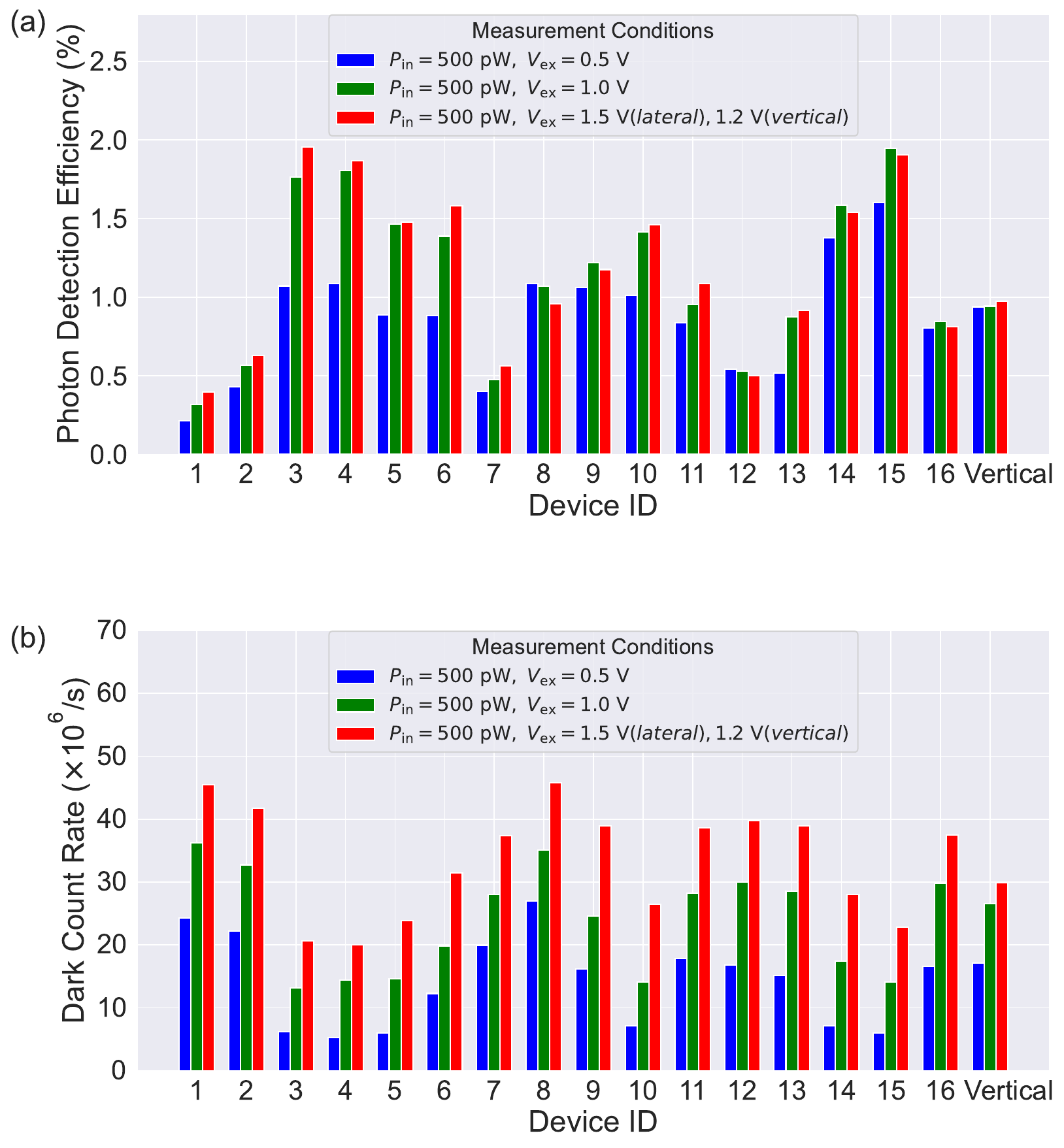}
    \caption{\label{fig:device_summary} 
    (a) Photon detection efficiency (PDE) and (b) Dark count rate (DCR) measurements for all SPAD variants 
    (16 lateral and one vertical) at three different excess bias voltages of 0.5\,V, 1.0\,V, and 1.5 V (1.2 V for vertical) at an input optical power of $P_\mathrm{in}$\,=\,500\,pW.
    The details of each device ID is listed in Table~\ref{tab:BV}.}
\end{figure}

\begin{table*}
\caption{\label{tab:BV}Device identifiers and their corresponding design parameters and breakdown voltages}
\begin{tabularx}{0.55\textwidth}{
  | >{\hsize=0.8\hsize\centering\arraybackslash}X
  | >{\centering\arraybackslash}X
  | >{\hsize=1.2\hsize\centering\arraybackslash}X
  | >{\centering\arraybackslash}X
  |
}
    \toprule
    Device ID & Intrinsic Region & Junction & Breakdown\\
    & Width $w$ ($\upmu$m) & Displacement $\Delta$ ($\upmu$m) & Voltage (V) \\
    \hline
    01 & 0 & 0 & 13.75 \\ 
    02 & 0 & 0.30 & 13.00 \\ 
    03 & 0 & 0.45 & 15.50 \\ 
    04 & 0 & 0.60 & 15.25 \\ \hline
    05 & 0.30 & 0 & 16.00 \\ 
    06 & 0.30 & 0.30 & 15.25 \\ 
    07 & 0.30 & 0.45 & 14.25 \\ 
    08 & 0.30 & 0.60 & 13.50 \\ \hline
    09 & 0.60 & 0 & 17.50 \\ 
    10 & 0.60 & 0.30 & 18.75 \\ 
    11 & 0.60 & 0.45 & 17.50 \\ 
    12 & 0.60 & 0.60 & 17.75 \\ \hline
    13 & 0.45 & 0 & 17.75 \\ 
    14 & 0.45 & 0.30 & 22.00 \\ 
    15 & 0.45 & 0.45 & 22.50 \\ 
    16 & 0.45 & 0.60 & 17.50 \\ \hline
    Vertical & - & - & 8.00 \\ \hline
\end{tabularx}
\end{table*}

\section*{References}
\bibliography{refs}